\documentclass[11pt]{article}

\usepackage[top = 30truemm, bottom = 30truemm, left=25truemm, right = 25truemm]{geometry}
\usepackage{setspace}
\usepackage{indentfirst}
\usepackage{amsmath, amsthm, amssymb}
\usepackage{bm}
\usepackage{graphicx}
\usepackage{cite}
\usepackage{url}

\title{Fractal Patterns in Spatial Distribution of Population by Age Group}
\date{}
\author{
Mariko I. Ito$^{1,\ast}$, and Takaaki Ohnishi$^{1}$\\
\\
\small{$^{1}$Graduate School of Artificial Intelligence and Science, Rikkyo University,} \\
\small{3-34-1, Nishi-Ikebukuro, Toshima-ku, Tokyo, Japan}\\
\small{$^{\ast}$ marikoito.rnu1@gmail.com}
}

\begin{document}
\maketitle

\doublespacing
\begin{abstract}
Multifractality is one of the patterns observed in spatial distributions of populations. 
In this study, we performed multifractal analysis for populations by age group in the capital area of Japan.
Each population group generally exhibited the property of densely populated locations being relatively connected to each other.
We also investigated the dependence of multifractality on the age group. 
Multifractal measures showed that the group consisting of young working-age people exhibits strong heterogeneity and concentration of population, while the population consisting of elder people exhibits a relatively homogeneous nature.
\end{abstract}


\section{Introduction}\label{sec:intro}

Multifractality has been observed in the spatial distribution of populations and various other artefacts, e.g., buildings and streets, in cities and regions across the world~\cite{ozikPRE,appleby1996multifractal,murcioPRE,HU2012161,ARIZAVILLAVERDE20131,chen2013multifractal,batty1995new,salat2018uncovering}.
Fractality is the nature where the mass in a region exhibits power law dependence on the size of the region. 
The exponent in the power-law relationship is called the fractal dimension~\cite{peitgen2006chaos}. 
An object can be regarded as exhibiting multifractality, if the local fractal dimension around each spot diverges~\cite{salat2018uncovering,chen2004multi,salat2017multifractal}. 

Ozik et al. (2005) suggested an explanation for the process of generating multifractality in a population~\cite{ozikPRE}.
They showed that the multifractal nature of the spatial distribution of a population can be generated by a process in which children settle around the locations of their parents stochastically.
In addition to this approach, we focus on the lifestyles and ages of the people in this study.
For example, people currently in their working-age may have stronger motivation to live in locations with good accessibility to the city centre or industrial clusters~\cite{diodato2018industries,o2019unravelling,schlapfer2014interface,arcaute2015interface} than elderly people who have retired.
Moreover, selective migration, where people migrate to more preferable places, is known to be a driving force of generating cities and cultures~\cite{ilya2014,BOYD2009331}.
It should be natural to assume that such lifestyle traits and migration of individuals can affect the spatial distribution of the population by age group.

In this study, we investigated multifractality in the spatial distribution of the population by age group. 
We evaluated the difference in spatial distributions of four age groups based on their multifractal properties. 
Our analysis showed significant heterogeneity in the population consisting of young working-age people, while the population of elderly people exhibit homogeneity in their multifractal properties.

This paper is organized as follows. 
We briefly introduce the concept of multifractals, and explain the method of our multifractal analysis in Section~\ref{sec:material}.
Results of the multifractal analysis and the interpretations of the multifractal measures are discussed in Section~\ref{sec:result}.
We discuss our results and future prospects in Section~\ref{sec:discussion}.

\section{Materials and methods}\label{sec:material}
\subsection{Data}

\begin{figure}[!h]
\centerline{\includegraphics[width=10cm]{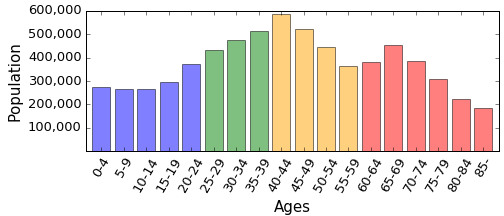}}
\caption{Histogram of the population by age. The population is classified into four groups according to age: Group A with ages 0 to 24 (blue bars), Group B with ages 25 to 39 (green bars), Group C with ages 40 to 59 (yellow bars), and Group D with ages $\geq$ 60 (red bars).}
\label{fig:hist}
\end{figure}

\begin{figure}[t]\vspace*{4pt}
\centerline{\includegraphics[width=10cm]{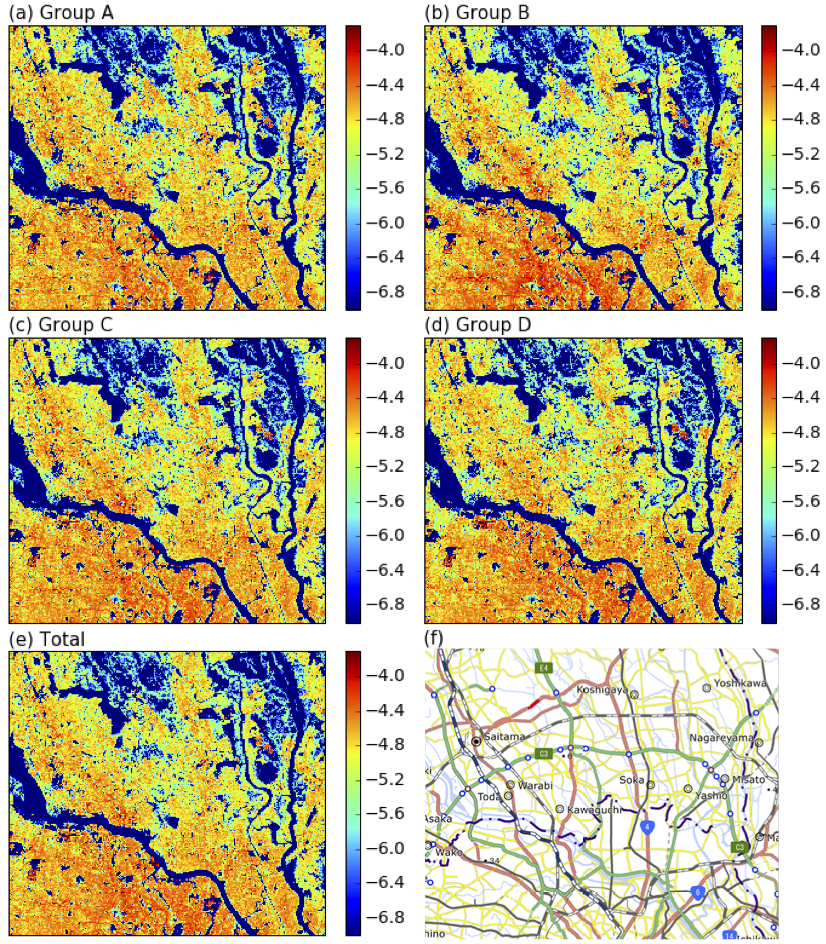}}
\caption{Spatial distribution of the population in (a)Group A, (b)Group B, (c)Group C, (d)Group D, (e)the total population (Total), (f)Analysed region. (1:1,000,000 INTERNATIONAL MAP, Geospatial Information Authority of Japan).
The heatmap shows a value of 
$\log_{10}$[(population in the mesh)/(population of the group) $+ 1.0\times 10^{-8}$] for each mesh.}
\label{fig:spatial_dist}
\end{figure}

We analysed data from the Japanese 100-Meter Estimated Mesh Data of the 2015 National Census\cite{zenrin}.
The data consists of estimated populations in each 100-meter mesh---hereafter referred to as `mesh'.
The exact size of a mesh is 3 seconds in the latitude direction and 4.5 seconds in the longitude direction.
The area concerned in this study is located northeast of the capital area in Japan, and it includes many commuter towns for the capital.
In the analysed area, the range of the latitude is from $35^\circ 42' 45''$ to $35^\circ 55' 30''$, and that of the longitude is from $139^\circ 36' 09''$ to $139^\circ 55' 16.5''$.
The analysed area contains $2^8\times 2^8 = 65,536$ meshes.
Fig.~\ref{fig:hist} is the histogram of the population by age in this area.
We classified the population into four groups according to their age: Group A with ages from 0 to 24, Group B with ages from 25 to 39, Group C with ages from 40 to 59, and Group D with ages greater than or equal to 60.
We determined the ages in each group in such a way that a group has a similar population to each other.
Total population and the population in each group are shown in Table~\ref{tab:data}.
In this paper, we call non-empty locations as {\it support}.
Table~\ref{tab:data} also shows the number of meshes and the maximum and mean populations in a mesh on the support.
The spatial distribution of the population in each group is shown with a heatmap in Fig.~\ref{fig:spatial_dist}.

\begin{table}[!h]
\caption{Data summary. 
The total population (Total) and the population in each age group (Group A, B, C and D) are shown in the second column.
The number of meshes on the support is shown in the third column. 
The maximum and the mean population in a mesh on the support are shown in the fourth and the fifth columns, respectively.}
\begin{tabular}{lrrrr}
\hline
Group & Population & \# Support & Max & Mean\\
\hline
Total & 6,745,402 & 51,836 & 1,966 & 130.130\\
Group A & 1,470,674 & 51,834 & 576 & 28.373\\
Group B & 1,419,504 & 51,831 & 569 & 27.387\\
Group C & 1,918,717 & 51,834 & 576 & 37.017\\
Group D & 1,936,506 & 51,834 & 935 & 37.360\\
\hline
\end{tabular}
\label{tab:data}
\end{table}

\subsection{Multifractal analysis}\label{subsec:analysis}

We begin by briefly introducing the concept of a multifractal based on the box-counting method.
As mentioned in Section~\ref{sec:intro}, an object can be regarded as exhibiting fractality if the mass in each region of the object, $m(\varepsilon)$, increases with the size $\varepsilon$ of the region according to a power law, as $m(\varepsilon)\sim \varepsilon^D$.
Multifractality is the nature where the exponent $D$ diverges in the object~\cite{meakin1998fractals,salat2018uncovering,salat2017multifractal,jiang2019multifractal}.
Let us assume that grid lines are drawn on the object, and it is covered by non-overlapping boxes with the same size $\varepsilon$ on one side that are defined by the grid lines.
The {\it probability measure} $P_{i,\varepsilon}$ on the $i$-th box can be defined as the ratio of the mass inside the box to the entire mass.
If $P_{i, \varepsilon}$ increases with $\varepsilon$ according to the power law for any $i$, as the following:
\begin{equation}
    P_{i,\varepsilon}\sim \varepsilon^\alpha, \varepsilon \rightarrow 0,
    \label{eq:alp_def}
\end{equation}
Then, this implies that fractality can be seen around each point of the object.
The exponent $\alpha$ can therefore be regarded as the local fractal dimension, and it is called the {\it singularity strength}. 
Subsequently, let $N(\varepsilon, \alpha)$ be the number of boxes that satisfy $P_{i,\varepsilon}\sim \varepsilon^{\tilde{\alpha}}$, where $\tilde{\alpha}\in [\alpha, \alpha+\Delta\alpha]$.
If $N(\varepsilon, \alpha)$ decreases with $\varepsilon$ as 
\begin{equation}
    N(\varepsilon, \alpha)\sim \rho(\alpha)\varepsilon^{-f(\alpha)}\Delta\alpha,
    \label{eq:spectrum_def}
\end{equation}
the object can be regarded as having a multifractal structure.
From Equation~(\ref{eq:spectrum_def}), the exponent $f(\alpha)$ is interpreted as the fractal dimension of the arrangement of points with singularity strength $\alpha$, and this dimension is called the {\it spectrum}.
Curves of $(\alpha, f(\alpha))$ are called {\it multifractal curves} hereafter.
The singularity strength $\alpha$ and the spectrum $f(\alpha)$ have a relationship with the {\it generalized dimension}~\cite{jiang2019multifractal,stanley1988multifractal}.
The $q$-th generalized dimension $D_q$ is defined as follows.
We first define $\tau_q$ as
\begin{equation}
    \tau_q = \lim_{\varepsilon\rightarrow 0}\frac{\log\sum_i P_{i, \varepsilon}^q}{\log\varepsilon}.
    \label{eq:tau_q}
\end{equation}
Then, $D_q$ for $q\neq 1$ is defined as
\begin{equation}
    D_q = \frac{1}{q-1}\tau_q.
    \label{eq:dq}
\end{equation}
When $q=1$, 
\begin{equation}
    D_1 = \lim_{\varepsilon\rightarrow 0}\frac{\sum_i P_{i, \varepsilon}\log P_{i, \varepsilon}}{\log\varepsilon}.
    \label{eq:dq_}
\end{equation}
In the summation in Equations~(\ref{eq:tau_q}, \ref{eq:dq_}), the $i$-th term is summed when the $i$-th box is not empty, i.e., $P_{i, \varepsilon}\neq 0$.

It is known that under some approximations, the singularity strength $\alpha_q$ and the spectrum $f(\alpha_q)$ can be derived from $\tau_q$ for each $q$ as the following:
\begin{eqnarray}
    & \alpha_q = \displaystyle\frac{d\tau_q}{dq},
    \label{eq:alpha_approx}\\
    & f(\alpha_q) = \alpha_q q - \tau_q.
    \label{eq:falpha_approx}
\end{eqnarray}

\begin{figure}[t]\vspace*{4pt}
\centerline{\includegraphics[width=10cm]{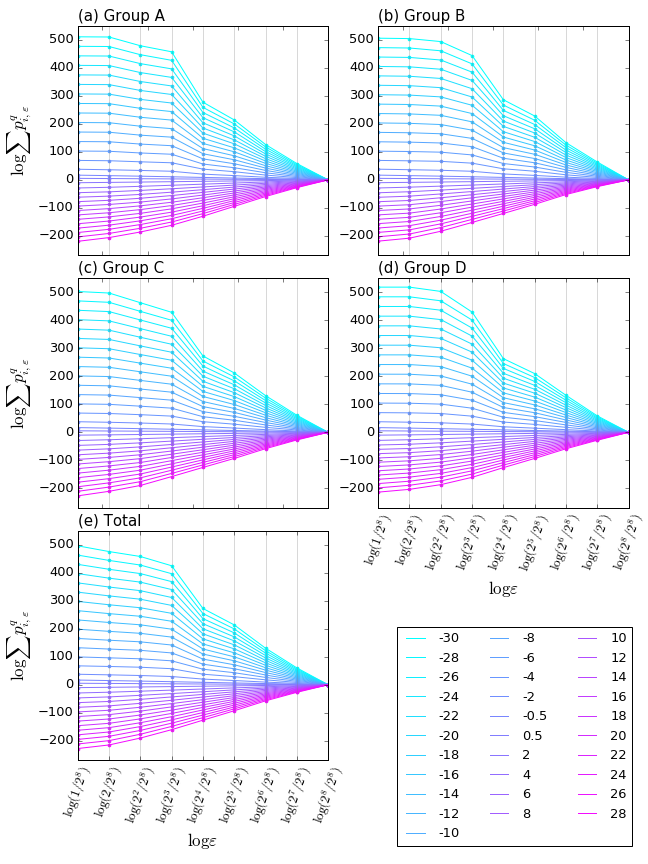}}
\caption{$\log\sum_i P_{i,\varepsilon}^q$ versus $\log\varepsilon$ for each $q$.
The colours of the plots correspond to the values of $q$ in the legends. Panels (a), (b), (c), (d), and (e) are for Groups A, B, C, D and the total population, respectively.}
\label{fig:loglog}
\end{figure}

We now explain the method of our multifractal analysis of the spatial distribution of the population, which is performed based on the box-counting method described above.
We assign the size of one side of a mesh to $\varepsilon = 1/2^8$.
Gridlines are drawn on the analysed region, and the region is covered by non-overlapping boxes of the same size that are defined by the grid lines.
As the sizes of boxes $\varepsilon$, we consider $\varepsilon = 1/2^8, ~2/2^8,~2^2/2^8,..., 2^7/2^8, 1$.
For example, when the region is covered by boxes of size $\varepsilon = 2/2^8$, one box contains $2\times 2$ meshes and there are $2^7\times 2^7$ boxes covering the region.
We define the probability measure $P_{i, \varepsilon}$ on the $i$-th box as the ratio of the population inside the $i$-th box to the entire population for each group, when the region is covered by boxes of size $\varepsilon$.
Unlike an ideal multifractal structure, our data does not have an infinitesimal structure.
The spatial distribution of the population (of each group) cannot exhibit multifractal property rigorously; therefore, we evaluate the range of $\varepsilon$ and $q$, where we can consider the spatial distribution as exhibiting multifractality.
We evaluated the range of $\varepsilon$ and $q$ for which the spatial distribution exhibits multifractality by examining the linearity of the relationship between $\log\sum_i P_{i,\varepsilon}^q$ and $\log\varepsilon$ based on the frequently used method~\cite{appleby1996multifractal,murcioPRE,chen2013multifractal,HU2012161,saa:hal-00302910,SUN2001553,TORRE2018167,grau2006comparison}. 
Plots of $\log\sum_i P_{i,\varepsilon}^q$ against $\log\varepsilon$ for each population group is shown in Fig.~\ref{fig:loglog}.
In our analysis, linearity is determined by examining whether or not the coefficient of determination of the linear regression of $\log\sum_i P_{i,\varepsilon}^q$ by $\log\varepsilon$ exceeds 0.99.
Consequently, we regard that multifractality can be observed in the spatial distribution of the population when $\varepsilon$ is more than $2^3$, for each group.
In addition, we set the range of $q$ from -30 to 30. 
We regard the slope of the linear regression of $\log\sum_i P_{i,\varepsilon}^q$ by $\log\varepsilon$ as $\tau_q$, based on Equation~(\ref{eq:tau_q}).
The generalized dimension $D_q$ is derived by Equations~(\ref{eq:dq}, \ref{eq:dq_}).
Subsequently, we derived the singularity strength $\alpha_q$ and the spectrum $f(\alpha_q)$ by linear regressions according to the following formulae:
\begin{eqnarray}
&\alpha_q = \displaystyle\lim_{\varepsilon\rightarrow 0}
\frac{\sum_i \mu_{i,\varepsilon,q} \log P_{i, \varepsilon}}
{\log\varepsilon} 
\label{eq:chhabra_a}\\
&f(\alpha_q) = \displaystyle\lim_{\varepsilon\rightarrow 0}
\frac{\sum_i \mu_{i,\varepsilon,q} \log \mu_{i,\varepsilon,q}}
{\log\varepsilon},
\label{eq:chhabra_f}
\end{eqnarray}
where $\mu_{i,\varepsilon,q} = P_{i, \varepsilon}^q / \sum_j P_{j, \varepsilon}^q$. 
These formulae are directly derived from Equations~(\ref{eq:tau_q}, \ref{eq:alpha_approx}, \ref{eq:falpha_approx}) and have often been used in previous studies~\cite{meakin1998fractals,PhysRevLett.62.1327,PhysRevA.40.5284}.

\section{Results}\label{sec:result}

\subsection{Generalized dimension}
\begin{figure}[t]\vspace*{4pt}
\centerline{\includegraphics[width=8cm]{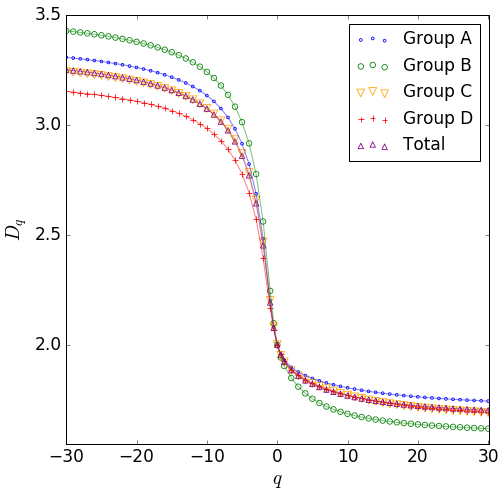}}
\caption{The $q$-th generalized dimension versus $q$.}
\label{fig:dq}
\end{figure}

In Fig.~\ref{fig:dq}, the $q$-th generalized dimension $D_q$ against $q$ is shown for the population in each age group and the total population.

The dependence of $D_q$ on $q$ varies in groups, and this can represent the strength of the heterogeneity in the spatial distribution as the following discussion.
As we can understand from Equation~(\ref{eq:tau_q}), a large value of $q$ corresponds to a great contribution of the boxes with large probability measures $P_{i,\varepsilon}$ to the sum.
Therefore, the boxes with high densities are significantly incorporated into the calculation of the $q$-th generalized dimension $D_q$ when the value of $q$ is large (see Equations~(\ref{eq:dq},~\ref{eq:dq_})).
On the other hand, boxes with small densities are emphasized in the calculation when $q$ is negative. 
Therefore, the value of $D_q$ for large (negative and small) $q$ reflects the nature of densely (sparsely) populated locations.
Thus, the extent to which $D_q$ declines with $q$ denotes the difference between the arrangement of densely populated locations and that of sparsely populated locations.

The generalized dimension $D_q$ of Group B declines with $q$ more rapidly than in other groups, which implicates the strong heterogeneity in the spatial distribution as mentioned above.
Meanwhile, Group D shows the smallest range of decline in $D_q$.
For negative values of $q$, $D_q$ for Group C exhibits a value close to that of the total population.
For positive values of $q$, the values of $D_q$ for Groups C and D are very close to that of the total population. 
Therefore, these results suggest that Group C (Groups C and D) exhibit a nature similar to the total population from the viewpoint of fractal geometry when we observe the sparsely (densely) populated locations.
The generalized dimension of Group A is relatively close to that of Group C and the total population. 
This may be because many young people in Group A (ages from 0 to 24) live with their parents in Group C.

\subsection{Multifractal curve}

\begin{figure}[t]\vspace*{4pt}
\centerline{\includegraphics[width=8cm]{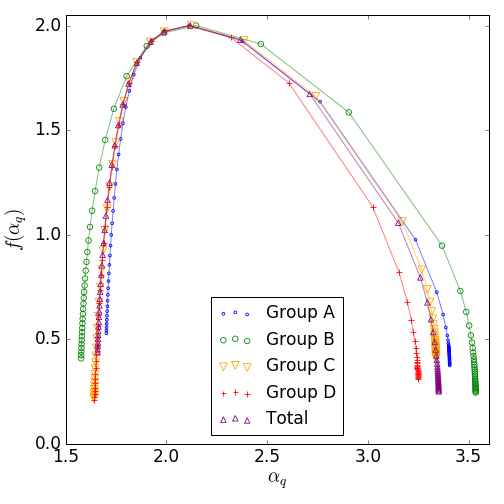}}
\caption{Multifractal curve. The horizontal axis is the singularity strength $\alpha_q$, and the vertical axis is the spectrum $f(\alpha_q)$.}
\label{fig:spectra}
\end{figure}

\begin{figure}[t]\vspace*{4pt}
\centerline{\includegraphics[width=12cm]{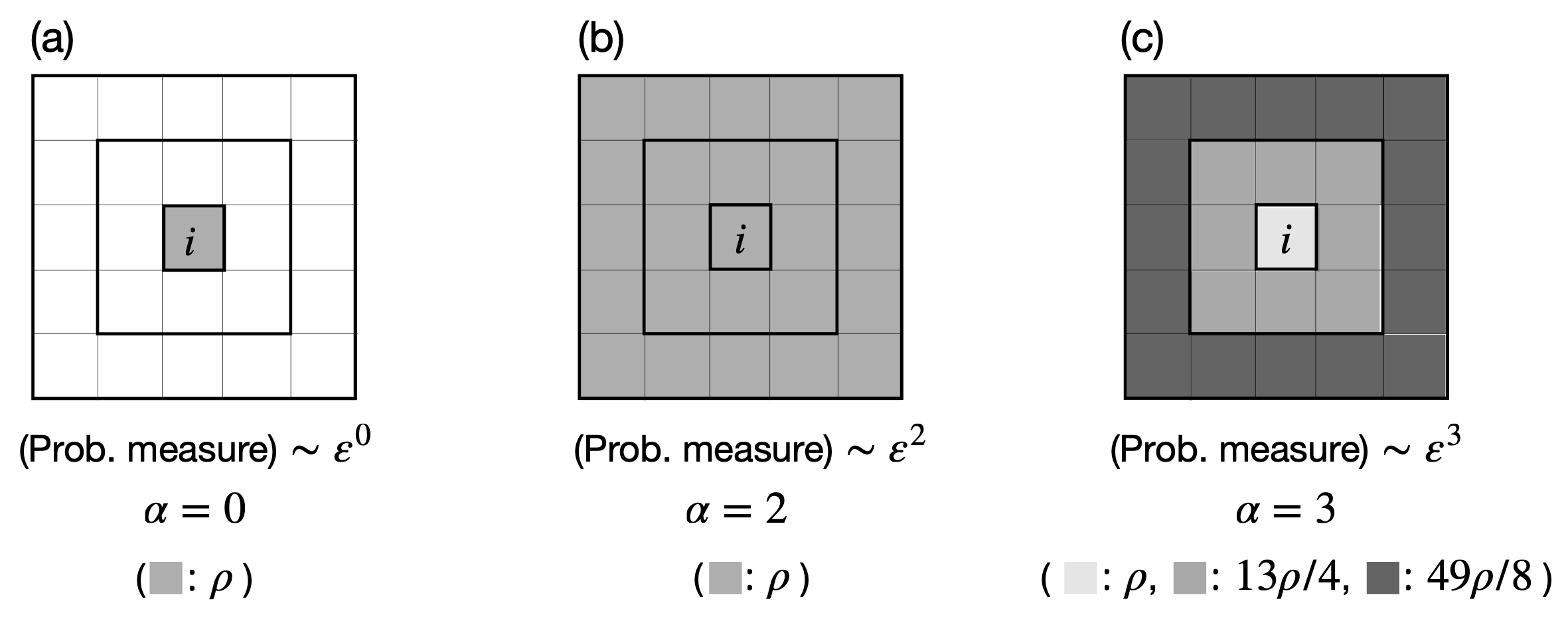}}
\caption{A schematic image of the singularity strength and the densities compared to the surroundings. 
A location corresponds to a smallest square. 
The gray-scale represents the probability measure on each location as shown in each panel. (a) Only the $i$-th location has non-zero density, and the other locations are empty. 
The probability measure on the box remains $\rho$, even when the size of the box $\varepsilon$ increases as emphasized by the thick line.
Therefore, the singularity strength of the $i$-th location is 0.
(b) All locations are filled with the same density.
The probability measures of the boxes are $\rho$, 9$\rho$ and 25$\rho$, for the smallest, the second smallest and the largest boxes (emphasized by the thick line), respectively.
Thus, the singularity strength of the $i$-th location is 2.
(c) The $i$-th location is sparse compared to the surroundings.
The probability measure of the boxes are $\rho$, 27$\rho$ and 125$\rho$, for the smallest, the second smallest and the largest boxes (emphasized by the thick line), respectively.
Thus, the singularity strength of the $i$-th location is 3.}
\label{fig:image}
\end{figure}

Multifractal curves for each group and the total population are shown in Fig.~\ref{fig:spectra}.
As mentioned in Section~\ref{subsec:analysis}, we derived the singularity strength $\alpha_q$ and spectrum $f(\alpha_q)$ pairs for each $q$.
The singularity strength $\alpha_q$ is generally smaller when the value of $q$ is smaller in a multifractal curve.
The spectrum $f(\alpha_q)$ generally takes the maximum value when $q=0$, in which the difference of densities in the boxes is not considered.
Divided at the mode where $q=0$, the left-hand and right-hand sides of the curve correspond to the positive and negative values of $q$, respectively.
This can be understood from the following explanation.
When we focus on a densely populated location such as in the $i$-th location in panel (a) of Fig.\ref{fig:image}, the probability measure on a region does not increase rapidly by expanding the region around the $i$-th location as represented by the thick lines. 
Thus, such locations have a low singularity strength, as defined by Equation~(\ref{eq:alp_def}).
Note that such densely populated locations are emphasized when $q$ is large.
On the other hand, around a sparsely populated location compared to the surroundings (the $i$-th location in Fig.\ref{fig:image}(c)), the probability measure on a region increases rapidly by expanding the size of the region, and the location exhibits a high singularity strength.
If the location has the same density as the surroundings (the $i$-th location in Fig.\ref{fig:image}(b)), the location exhibits a singularity strength of 2.

In Fig.~\ref{fig:spectra}, all multifractal curves are skewed, and the mode of each is biased to the left.
The range of $\alpha_q$ in each multifractal curve is narrower when $q$ is positive (left-hand side) than when $q$ is negative (right-hand side), and $\alpha_q$ does not decline significantly for positive $q$.
Therefore, it can be inferred that the population does not concentrate to isolated spots, but the densely populated areas are connected to each other, while sparsely populated locations are isolated.

Subsequently, we compare the multifractal curves.
The multifractal curve of Group B exhibits the widest range of $\alpha_q$.
Therefore, in the spatial distribution of Group B, there should be both significantly dense and significantly sparse locations compared to the surroundings, and such heterogeneity is stronger than that of the other groups.
Meanwhile, Group D exhibits the narrowest range of $\alpha_q$.
Thus, $\alpha_q$, that corresponds to the density compared to the surroundings, varies less in Group D.

Groups C and D exhibit curves similar to each other for positive values of $q$ (the left-hand side).
However, these curves are separated for negative values of $q$ (the right-hand side).
Therefore, it is inferred that the nature of the spatial distribution of Group C is similar to that of Group D when we limit our observation to the densely populated area.
In addition, the multifractal curve of the total population exhibits an overlap with that of Group C in a large range.

Recall that the $q$-th generalized dimensions of Group C and the total population are close to each other for negative values of $q$, and that of Groups C and D, and the total population are close to each other for positive values of $q$.
Such a similarity relationship of the generalized dimensions among the age groups is roughly consistent with that of the multifractal curves.
However, the spectra $f(\alpha_q)$ of Groups C and D take significantly different values from that of the total population.
Unlike the total population, $f(\alpha_q)$ of Groups C and D drops to a large extent as $q$ increases, i.e., as $\alpha_q$ decreases, when $q$ is positive.
Recall that $f(\alpha_q)$ stands for the fractal dimension of the arrangement of locations exhibiting a singularity strength $\alpha_q$.
Thus, we can infer that the arrangement of locations with a certain singularity strength, i.e., a relative density against the surroundings, rapidly loses the fractal dimension as the singularity strength decreases in Groups C and D. 

\section{Discussion}\label{sec:discussion}

We investigated multifractality in the spatial distribution of a population focusing on the age difference.
In our analysis, the population is divided into four groups according to the age: Group A, B, C and D, from younger to older.

Results of the multifractal analysis showed that Group D exhibits homogeneous nature, while Group B showed the strongest heterogeneity in the spatial distribution from the viewpoint of fractal geometry.
The multifractal curve of Group B has a lower singularity strength than that of the other groups, and such a small singularity strength implies the existence of locations with significantly high densities compared to the surroundings.
Therefore, the concentration of the population should be stronger in Group B than in the other groups.
However, even in the multifractal spectrum of Group B, the singularity strength is more than 1 and is not extremely small.
This implies that strong concentrations such as in Group B can be rarely observed in isolated spots, but can be observed in connected areas, e.g., along railway tracks that have a one-dimensional shape.
Group B consists of ages 25 to 39, and they are in the working-age.
One possibility is that the concentration of the population occurs around the locations where they can take advantage of convenient transportations to their working places.
Meanwhile, Groups C and D exhibit similar multifractal properties when $q$ is positive, i.e., in the densely populated area.
However, when $q$ is negative, $D_q$ and $\alpha_q$ of Group D are less than that of Group C.
We can infer that the nature of densely populated locations of Group C is qualitatively similar to that of Group D, but Group D exhibits more homogeneous spatial distribution than the others when we expand our view to include the sparsely populated area.

Our analysis showed that the multifractal properties in the spatial distribution of a population depend significantly on the ages of the people, while they are distributed on the same geographical substrate. 
Such a difference in the multifractal properties by age suggests that not only single settlements, but also changes of their residence according to their ages and lifestyles should significantly contribute to the process of generating the multifractal patterns in the spatial distributions of the population. 
In addition, we should consider external factors, e.g., the developments of land and estate, to be included in such a process. For example, development of a convenient area can attract young people who consider purchasing their residences, and it can generate a concentrated population, which is a multifractal pattern specifically observed in Group B.

We also expect that the temporal development of the multifractal nature in the spatial distribution of a population can be affected by the age of the people.
Previous studies showed that spatial distributions associated with city morphology, e.g., buildings and streets, have temporally developed to the packed state with the strong homogeneity~\cite{encarnaccao2012fractal,murcioPRE}.
It should be interesting to see whether the whole population also develops towards a homogeneous nature.
If so, by revealing which ages contribute more/less to such development to homogeneity, we will be able to enhance our understanding of the development of cities. 

\section{Conclusion}
We investigated the multifractal properties of the spatial distribution of a population by age group in the Japanese capital area.
Populations in each age group exhibited different multifractal natures.
The population consisting of young working-age people in particular exhibited the strongest heterogeneity in multifractal measures, implying a strong concentration of population in some locations.

\vskip1pc

\noindent
{\bf Acknowledgements. }This work was supported by JSPS KAKENHI Grant Number 19K21578.

\pagebreak

\bibliographystyle{vancouver}
\bibliography{reference}

\begin{thebibliography}{10}

\bibitem{ozikPRE}
Ozik J, Hunt BR, Ott E.
\newblock Formation of multifractal population patterns from reproductive
  growth and local resettlement.
\newblock Physical Review E. 2005;72:046213.

\bibitem{appleby1996multifractal}
Appleby S.
\newblock Multifractal Characterization of the Distribution Pattern of the
  Human Population.
\newblock Geographical Analysis. 1996;28(2):147--160.

\bibitem{murcioPRE}
Murcio R, Masucci AP, Arcaute E, Batty M.
\newblock Multifractal to monofractal evolution of the London street network.
\newblock Physical Review E. 2015;92:062130.

\bibitem{HU2012161}
Hu S, Cheng Q, Wang L, Xie S.
\newblock Multifractal characterization of urban residential land price in
  space and time.
\newblock Applied Geography. 2012;34:161--170.

\bibitem{ARIZAVILLAVERDE20131}
Ariza-Villaverde AB, Jiménez-Hornero FJ, Ravé EGD.
\newblock Multifractal analysis of axial maps applied to the study of urban
  morphology.
\newblock Computers, Environment and Urban Systems. 2013;38:1--10.

\bibitem{chen2013multifractal}
Chen Y, Wang J.
\newblock Multifractal characterization of urban form and growth: the case of
  Beijing.
\newblock Environment and Planning B: Planning and Design. 2013;40(5):884--904.

\bibitem{batty1995new}
Batty M.
\newblock New ways of looking at cities.
\newblock Nature. 1995;377(6550):574--574.

\bibitem{salat2018uncovering}
Salat H, Murcio R, Yano K, Arcaute E.
\newblock Uncovering inequality through multifractality of land prices: 1912
  and contemporary Kyoto.
\newblock PLOS ONE. 2018;13(4):1--19.

\bibitem{peitgen2006chaos}
Peitgen HO, J{\"u}rgens H, Saupe D.
\newblock Chaos and Fractals: New Frontiers of Science.
\newblock New York: Springer Science \& Business Media; 2006.

\bibitem{chen2004multi}
Chen Y, Zhou Y.
\newblock Multi-fractal measures of city-size distributions based on the
  three-parameter Zipf model.
\newblock Chaos, Solitons \& Fractals. 2004;22(4):793--805.

\bibitem{salat2017multifractal}
Salat H, Murcio R, Arcaute E.
\newblock Multifractal methodology.
\newblock Physica A: Statistical Mechanics and its Applications.
  2017;473:467--487.

\bibitem{diodato2018industries}
Diodato D, Neffke F, O’Clery N.
\newblock Why do Industries Coagglomerate? How Marshallian Externalities Differ
  by Industry and Have Evolved Over Time.
\newblock Journal of Urban Economics. 2018;106:1--26.

\bibitem{o2019unravelling}
O'Clery N, Heroy S, Hulot F, Beguerisse-Diaz M.
\newblock Unravelling the forces underlying urban industrial agglomeration.
\newblock arXiv preprint arXiv:190309279. 2019;.

\bibitem{schlapfer2014interface}
Schl{\"a}pfer M, Bettencourt LM, Grauwin S, Raschke M, Claxton R, Smoreda Z,
  et~al.
\newblock The scaling of human interactions with city size.
\newblock Journal of the Royal Society Interface. 2014;11(98):20130789.

\bibitem{arcaute2015interface}
Arcaute E, Hatna E, Ferguson P, Youn H, Johansson A, Batty M.
\newblock Constructing cities, deconstructing scaling laws.
\newblock Journal of The Royal Society Interface. 2015;12(102):20140745.

\bibitem{ilya2014}
Somin I.
\newblock FOOT VOTING, FEDERALISM, AND POLITICAL FREEDOM.
\newblock Nomos. 2014;55:83--119.

\bibitem{BOYD2009331}
Boyd R, Richerson PJ.
\newblock Voting with your feet: Payoff biased migration and the evolution of
  group beneficial behavior.
\newblock Journal of Theoretical Biology. 2009;257(2):331--339.

\bibitem{zenrin}
Zenrin Marketing Solutions Co.,Ltd.;.
\newblock Available from:
  \url{https://www.zenrin-ms.co.jp/gis_marketing/database/statistics/census_100m_mesh/}.

\bibitem{meakin1998fractals}
Meakin P.
\newblock Fractals, Scaling and Growth Far from Equilibrium. vol.~5.
\newblock Cambridge: Cambridge University Press; 1998.

\bibitem{jiang2019multifractal}
Jiang ZQ, Xie WJ, Zhou WX, Sornette D.
\newblock Multifractal analysis of financial markets: a review.
\newblock Reports on Progress in Physics. 2019;82(12):125901.

\bibitem{stanley1988multifractal}
Stanley HE, Meakin P.
\newblock Multifractal phenomena in physics and chemistry.
\newblock Nature. 1988;335(6189):405--409.

\bibitem{saa:hal-00302910}
Saa A, Gasc{\'o} G, Grau JB, Ant{\'o}n JM, Tarquis AM.
\newblock {Comparison of gliding box and box-counting methods in river network
  analysis}.
\newblock {Nonlinear Processes in Geophysics}. 2007;14(5):603--613.

\bibitem{SUN2001553}
Sun X, Chen H, Wu Z, Yuan Y.
\newblock Multifractal analysis of Hang Seng index in Hong Kong stock market.
\newblock Physica A: Statistical Mechanics and its Applications.
  2001;291(1):553--562.

\bibitem{TORRE2018167}
Torre IG, Losada JC, Heck RJ, Tarquis AM.
\newblock Multifractal analysis of 3D images of tillage soil.
\newblock Geoderma. 2018;311:167--174.

\bibitem{grau2006comparison}
Grau J, Méndez V, Tarquis AM, Díaz MC, Saa A.
\newblock Comparison of gliding box and box-counting methods in soil image
  analysis.
\newblock Geoderma. 2006;134(3):349--359.

\bibitem{PhysRevLett.62.1327}
Chhabra A, Jensen RV.
\newblock Direct determination of the f(\ensuremath{\alpha}) singularity
  spectrum.
\newblock Physical Review Letters. 1989;62:1327--1330.

\bibitem{PhysRevA.40.5284}
Chhabra AB, Meneveau C, Jensen RV, Sreenivasan KR.
\newblock Direct determination of the f(\ensuremath{\alpha}) singularity
  spectrum and its application to fully developed turbulence.
\newblock Physical Review A. 1989;40:5284--5294.

\bibitem{encarnaccao2012fractal}
Encarna{\c{c}}{\~a}o S, Gaudiano M, Santos FC, Tened{\'o}rio JA, Pacheco JM.
\newblock Fractal cartography of urban areas.
\newblock Scientific Reports. 2012;2:527.

\end{thebibliography}

\end{document}